\documentclass[useAMS,usenatbib]{mn2e}
\usepackage{natbib}
\usepackage{gensymb}
\usepackage{amssymb}
\usepackage{amsmath}
\usepackage{wasysym}
\usepackage{graphicx}
\usepackage{color}
\usepackage{textcomp}
\usepackage{mathrsfs}
\usepackage{subfigure}
\usepackage{tabularx}

\DeclareMathSymbol{\varOmega}{\mathord}{letters}{"0A}
\DeclareMathSymbol{\varSigma}{\mathord}{letters}{"06}

\DeclareMathSymbol{\varPsi}{\mathord}{letters}{"09}

\newcommand{\lsim}{\mathrel{\rlap{\lower4pt\hbox{\hskip1pt$\sim$}}
   \raise1pt\hbox{$<$}}}                % less than or approx. symbol
\newcommand{\gsim}{\mathrel{\rlap{\lower4pt\hbox{\hskip1pt$\sim$}}
   \raise1pt\hbox{$>$}}}                % greater than or approx. symbol
\begin{document}
\title[Clearing of debris]{The unseen planets of double belt debris disk systems}
\author[Shannon et al.]{Andrew Shannon$^1$, Amy Bonsor$^1$, Quentin Kral$^1$, \& Elisabeth Matthews$^2$ \\
$^1$Institute of Astronomy, University of Cambridge, Madingley Road, Cambridge, UK, CB3 0HA \\
$^2$Astrophysics Group, University of Exeter, Physics Building, Stocker Road, Exeter, UK, EX4 4QL}
\maketitle

\begin{abstract}
The gap between two component debris disks is often taken to be carved by intervening planets scattering away the remnant planetesimals.  We employ $N$-body simulations to determine how the time needed to clear the gap depends on the location of the gap and the mass of the planets.  We invert this relation, and provide an equation for the minimum planet mass, and another for the expected number of such planets, that must be present to produce an observed gap for a star of a given age.  We show how this can be combined with upper limits on the planetary system from direct imaging non-detections (such as with GPI or SPHERE) to produce approximate knowledge of the planetary system.

\end{abstract}


\section{Introduction}
Debris disks are circumstellar dust disks, produced by the destructive collisions of planetesimals leftover from the planet formation process \citep{2008ARA&A..46..339W}.  There exist a significant number of debris disks with two temperature components \citep{2008ApJ...677..630H}.  Modelling suggests that in at least a significant fraction of cases, these two temperature disks harbour two concentric debris rings, with a significant gap between them \citep{2014MNRAS.444.3164K} - somewhat analogous to the asteroid and Kuiper belts of the Solar system.

Also by analogy with the Solar system, the gap is often inferred to have been opened by planets scattering away the remnant planetesimals.  \citet{2007MNRAS.382.1823F} modelled the gap clearing as caused by multi-planet instabilities \citep{1996Icar..119..261C} producing `Nice model' like clearing of massive planetesimal belts \citep{2005Natur.435..466G}.  However, attempts to match such instabilities to observed debris disks suggest they must be rare events overall \citep{2009MNRAS.399..385B}, and thus they are unlikely to be the principle mechanism for gap clearing.  This rarity should also apply to the formation of a double ring by a single, eccentric, dynamically unstable planet, as modelled by \citet{2015MNRAS.453.3329P}.

\begin{figure}
  \centering
  %\subfloat
  {\includegraphics[width=0.49\textwidth,trim = 0 0 0 0, clip]{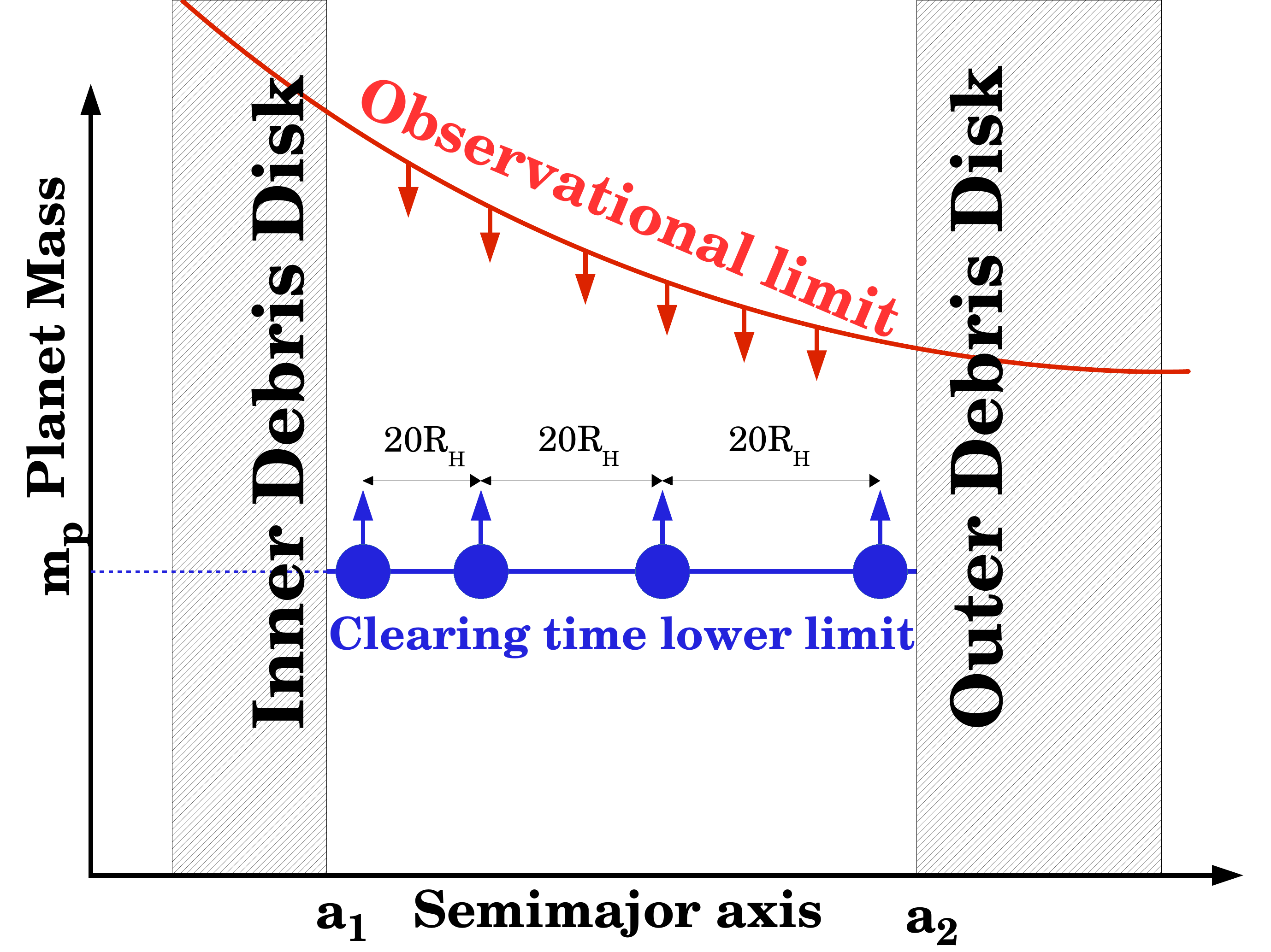}}
  %\subfloat
  \caption{Schematic representation of how the lower limit presented here produces an overall bounded view of the possible planetary system when combined with the upper limits from direct imaging non-detection.}
\label{fig:schematic}
\end{figure}

The time for a single planet to clear its chaotic zone was considered by \citet{2015ApJ...799...41M} and \citet{2015ApJ...798...83N}.  In the case of the observed gaps opened in double debris disk systems, the necessary planet mass is often too large to have escaped detection by direct imaging attempts.  This led \citet{2014IAUS..299..318S} to suggest that the observed gaps may be opened by several planets scattering away the remnant planetesimals.  Despite some attempts \citep{2007ApJ...666..423Z,2011MNRAS.418.1043Q,2011ApJ...735..109W,2011ApJ...739...31L}, a general theory of the stability of many-planet systems has not yet been developed.  Great success, however, has been enjoyed by $N$-body simulations \citep{1996Icar..119..261C,2009Icar..201..381S,2014MNRAS.437.3727K,2015ApJ...807...44P}.  Thus, to consider the case where gaps in double debris disks are caused by multiple planets scattering away the planetesimals leftover from the planet formation epoch, we use $N$-body simulations to calculate the clearing time for a given planetary system.  By inverting this relation, we recover an equation for the minimal planetary system that must be present in a gap for a system of a given age (figure \ref{fig:schematic}).

\section{Simulations}
To fill a gap that extends from $a_1$~to $a_2$~with $N$~planets spaced by $K$~mutual hill radii $\left(R_H\right)$, the planets must have mass
\begin{equation}
 \mu = \left(\frac{m_p}{m_*}\right) = \frac{12}{K^3} \frac{\left[\left(\frac{a_2}{a_1}\right)^{\frac{1}{N-1}}-1\right]^3}{\left[\left(\frac{a_2}{a_1}\right)^{\frac{1}{N-1}}+1\right]^3} .
 \label{eq:pmass}
\end{equation}
At the inner and outer edges of the gap, the chaotic zone will slightly widen the gap \citep{1980AJ.....85.1122W}, but this can be a rather nuanced problem \citep[e.g.,][]{1989Icar...82..402D,2009ApJ...693..734C,2012MNRAS.419.3074M,2015MNRAS.448..684S}.  As this zone is small compared to the inter-planet spacing, we neglect it for this simple model.  \citet{2013ApJ...767..115F} showed that the typical separation between planets in Kepler multi-planet systems is $21.7 \pm 9.5 R_{H}$.  We thus adopt $K = 20$~for our typical separation, and space the planets evenly.  Planets were given eccentricity distributed randomly and linearly from $e = 0$~to $e = 0.02$ \citep[roughly the Kepler multi-planet value,][]{2014ApJ...787...80H}, and inclinations distributed randomly and linearly from $ i = 0\degree$~to $i = 2\degree$~\citep[again, following Kepler multi-planet systems,][]{2014ApJ...790..146F}.  All planets are assigned a density of $\rho = 4~\rm{g}~\rm{cm}^{-3}$.  We place 100 test particles evenly between $a_1$~and $a_2$, with eccentricities from $e = 0$~to $e = 0.1$~and inclinations from $ i = 0\degree$~to $i = 10\degree$.

\begin{figure}
  \centering
  %\subfloat
  {\includegraphics[width=0.49\textwidth,trim = 150 50 100 50, clip]{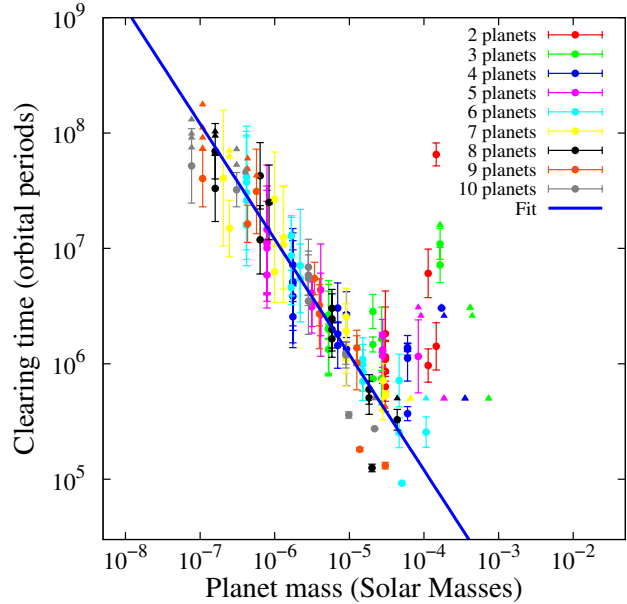}}
  %\subfloat
  \caption{Clearing time in orbital periods at the outermost edge of the gap, for all the systems.  The fit (equation \ref{eq:formfit}) matches most of the data well, although it fails in some cases with small numbers of  planets in wide gaps.}
\label{fig:result}
\end{figure}

We define the clearing time $\tau_{\rm{clear}}$~as the time it takes for half of the initial particles to no longer have a star-particle separation of between $a_1$~and $a_2$, whether they collide with a planet, the star, or are scattered or ejected from the belt.  In a few cases (which all failed equation \ref{eq:minnum}), we cut off simulations after $5 \times 10^8$~or more orbits at $a_1$; those are represented as lower limits in figure \ref{fig:result}, and not used in the fit for equation \ref{eq:minmass}.  As particles scattered to eccentric orbits move in and out of the $a_1 \rightarrow a_2$~belt, we take the first and last time 50 particles reside in the belt as our uncertainty.   Simulations were performed with MERCURY \citep{1999MNRAS.304..793C}.  The values of $a_1$~and $a_2$~we simulated are listed in table \ref{tab:sims}.

\begin{center}
\begin{table}

 \begin{tabularx}{0.47\textwidth}{| r || @{\extracolsep{\fill}} r r r r r r |}
  \hline
  $a_1$ & $a_2$ & $a_2$ & $a_2$ & $a_2$ & $a_2$ & $a_2$ \\
  \hline 
  1 au & 2 au & 3 au & 10 au & 30 au & 100 au &  \\
  3 au & 6 au &   & 10 au & 30 au & 100 au & 300 au \\
  10 au & 20 au &  &   & 30 au & 100 au & 300 au \\
  30 au & 60 au &  &   &    & 100 au & 300 au \\
  100 au & 200 au &  & &    &     & 300 au \\
  \hline
 \end{tabularx}
  \caption{The inner and outer edges of the belts that we simulated.  Each case was performed for $N = 2$~to $N = 10$~planets.}
  \label{tab:sims}

\end{table}
\end{center}

\begin{figure}
  \centering
  {\includegraphics[width=0.49\textwidth,trim = 150 50 100 50, clip]{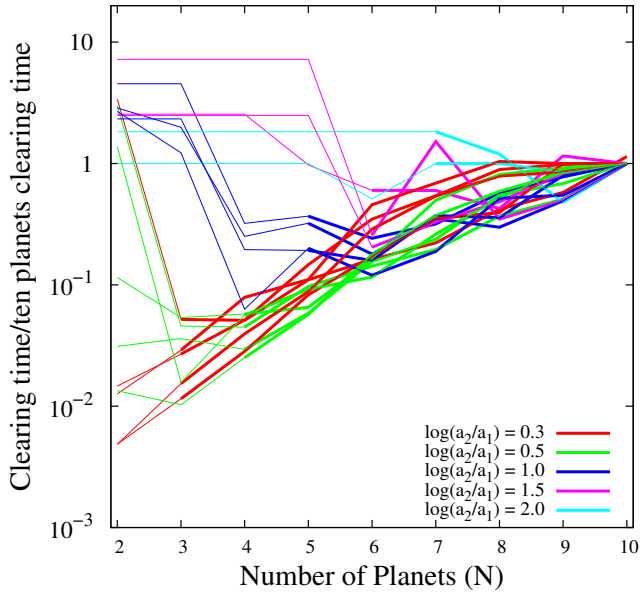}}
  \caption{The ratio of the clearing time for the $N = 10$~planet case, to the $N$~planet case, for each parameter choice listed in table \ref{tab:sims}.  Colours indicate $\log{\frac{a_2}{a_1}}$.  Simulations exhibit a general trend of decreasing clearing time with decreasing $N$ (and thus, increasing $m_p$) for systems that obey equation \ref{eq:minnum} (thick lines), but this can break and reverse for those that do not (thin lines).}
\label{fig:failure}
\end{figure}

Examining the data (figure \ref{fig:result}), we notice that in most cases, as the number of planets increased and their mass decreased for a given belt, the clearing time lengthened.  This trend held for systems as long as
\begin{equation}
 \frac{N}{2}-1 > \log{\frac{a_2}{a_1}} ,
 \label{eq:minnum}
\end{equation}
but for cases with wide belts and few planets, this trend can flatten or reverse (figure \ref{fig:failure}).  We exclude those cases when fitting the clearing times.  

\begin{figure}
  \centering
  %\subfloat
  {\includegraphics[width=0.49\textwidth,trim = 150 50 100 50, clip]{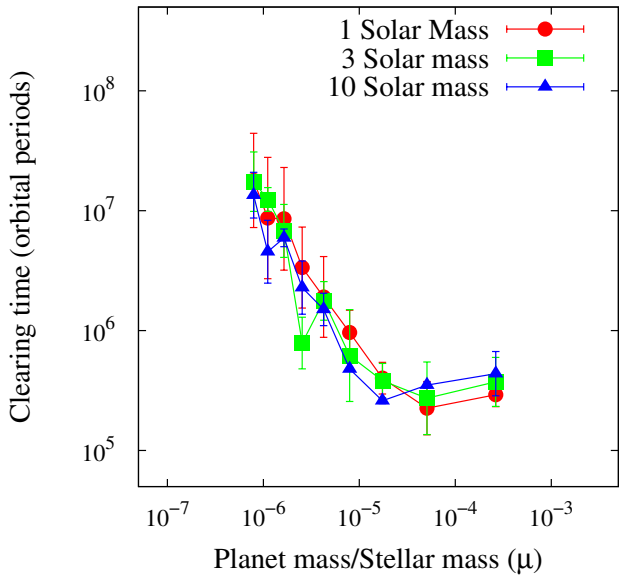}}
  %\subfloat
  \caption{Clearing time in orbital periods at the outermost edge of the gap, for $a_1 = 3~\rm{au}$~and $a_2 = 10~\rm{au}$, for $M_*~= 1~M_{\odot}, 3~M_{\odot},$~and $10~M_{\odot}$.  These simulations were done with $K=16$.  This shows that the clearing time scales with the stellar mass in the same manner as the secular time, i.e., $\tau \propto \sqrt{M_*}$.}
\label{fig:stellarmass}
\end{figure}

Fitting the simulation results with a power law of the form 
\begin{equation}
 \tau_{\rm{clear}} = \alpha \left(\frac{a_2}{\rm{1 au}}\right)^{\beta} \left(\frac{m_p}{m_\oplus}\right)^{\gamma} ,
 \label{eq:formfit}
\end{equation}
we find $\alpha = 4 \pm 1 \times 10^6~\rm{yrs}$, $\beta = 1.6 \pm 0.05$~and $\gamma = -0.94 \pm 0.04$
(figure \ref{fig:result}).  Simulations with $K = 16$~but otherwise the same parameters gave $\alpha = 2 \pm 0.2 \times 10^6$~but otherwise the same results, therefore we infer this method is not strongly sensitive to the exact choice of $K$.  After fitting the $a_2$~and $m_p$~dependence, $\tau_{\rm{clear}}$~has no further dependence on $a_1$~nor $N$. Equation \ref{eq:formfit} has the same scaling as, and is comparable in magnitude to, the secular interaction time for two equal-mass planets on nearby orbits\footnote{\citet{1999ssd..book.....M} exercise question 7.1}.  Thus we posit secular resonances may be key to clearing the test particles - and correspondingly, with few planets, the resonances are too sparse to cover the gap sufficiently to clear away most particles, resulting in the breakdown for systems that fail equation \ref{eq:minnum}.  As such, we set the scaling with stellar mass as it is for secular interactions.  We verify that this scaling corrects for stellar mass in figure \ref{fig:stellarmass}.

This result can be applied to observed double debris disk systems to infer the minimum planetary system that should be present in the gap.  For a star of age $\tau$, the minimum mass of the planets in the gap is
\begin{equation}
  m_p = \left(\frac{4 \rm{Myrs}}{\tau}\right) \left(\frac{a_2}{1~\rm{au}}\right)^{\frac{3}{2}}\left(\frac{M_*}{M_{\odot}}\right)^{\frac{1}{2}} m_\oplus ,
% \tau_{\rm{clear}} \approx 2 \times 10^6 \left(\frac{a}{\rm{1 au}}\right)^{1.5} \left(\frac{m_p}{m_\oplus}\right)^{-1} .
 \label{eq:minmass}
\end{equation}
and assuming typical spacing, the number of planets in the gap is
\begin{equation}
 N = 1 + \frac{\log{\left(\frac{a_2}{a_1}\right)}}{\log{\left(\frac{1 + 0.13\left(\frac{m_p}{m_\oplus}\right)^{\frac{1}{3}}\left(\frac{M_{\odot}}{M_*}\right)^{\frac{1}{3}}}{1 - 0.13\left(\frac{m_p}{m_\oplus}\right)^{\frac{1}{3}}\left(\frac{M_{\odot}}{M_*}\right)^{\frac{1}{3}}}\right)}} \ .
 \label{eq:numberofplanets}
\end{equation}

Equation \ref{eq:numberofplanets} can also be applied to the upper mass limit derived from observational non-detection to envision the maximal planetary system and thus produce a complete picture of what planetary systems could lie within the system (figure \ref{fig:schematic}).

There is some scatter about the relation but the only systematic trend is the breakdown when the mass of planets is large, corresponding to number of planets being small  (i.e., for systems that do not obey equation \ref{eq:minnum}).  This result is for equal mass planets with spacings typical of extra-solar systems; one might expect unusually compact or sparse systems to clear faster or slower.  For unequal masses, the situation is likely to be more complicated, but for the application considered here, it is reasonable to assume that the clearing will proceed no faster than equation \ref{eq:formfit} for the most massive planet.  For unequal spacings, \citet{2015ApJ...807...44P} showed that the {\em first} instability occurred sooner than for equal spacings.  However, here we consider the case of the average instability time for a very large number ($10^{6} \sim 10^{12}$) of planetesimals, and so expect the difference between equal and unequal spacings to be minimal.

\section{Real systems}

\subsection{Validation}

\subsubsection{A young system: HR 8799}
HR 8799 is orbited by four giant planets \citep{2008Sci...322.1348M,2010Natur.468.1080M}, nestled snugly between two debris disks \citep{2009ApJ...705..314S}, making it an ideal case for evaluating this model.  The outer edge of the cleared zone is $\sim 145~\rm{au}$~\citep{2016MNRAS.tmpL..24B}.  The most common age estimate is $\sim 30~\rm{Myrs}$~\citep{2010lyot.confE..42D,2011ApJ...732...61Z,2012ApJ...761...57B}, and the stellar mass is $M_* \sim 1.5~M_{\odot}$ \citep{1999AJ....118.2993G}.  From equation \ref{eq:minmass}, this requires a minimum planet mass of $\sim~285~m_{\oplus}$~to have cleared the gap between the two belts.  With the inner belt extending to $\sim 15~\rm{au}$, this mass requires $\sim 2.3$~planets to fill the gap (equation \ref{eq:numberofplanets}).  Since we require an integer number of planets, the minimal planetary system is three $285~m_\oplus$~planets, packed more tightly than average.  This is about a factor of $\sim 10$~less than the best estimates of the masses of the four planets \citep{2011ApJ...729..128C}, so our inferred lower limit is indeed compatible with the real system, whose unusual compactness is due to its special dynamic state \citep{2010ApJ...710.1408F,2014MNRAS.440.3140G}.

\subsubsection{An old system: Solar System}

The Solar system has a gap between its two debris disks which extends from the asteroid belt at about 3.5 au to the Kuiper belt at around 39 au.  With an age of $4.56 \times 10^9$~years, the minimum planet mass needed to clear the gap in our system is $\sim 0.2~m_\oplus$, roughly twice the mass of Mars, about seventeen of which would fit between the two belts.  The four planets observed between the asteroid and Kuiper belt are indeed more massive than this minimum.  Thus, we show applying our model to both the young system HR 8799, and the old Solar system, the only two systems for which we have good data on multiple planets between two debris disks, we recover a minimum planet mass that is less than the observed planet masses.

\subsection{Example Future observations}

\subsubsection{HD 38206}

\begin{figure}
  \centering
  %\subfloat
  {\includegraphics[width=0.48\textwidth,trim = 0 0 0 0, clip]{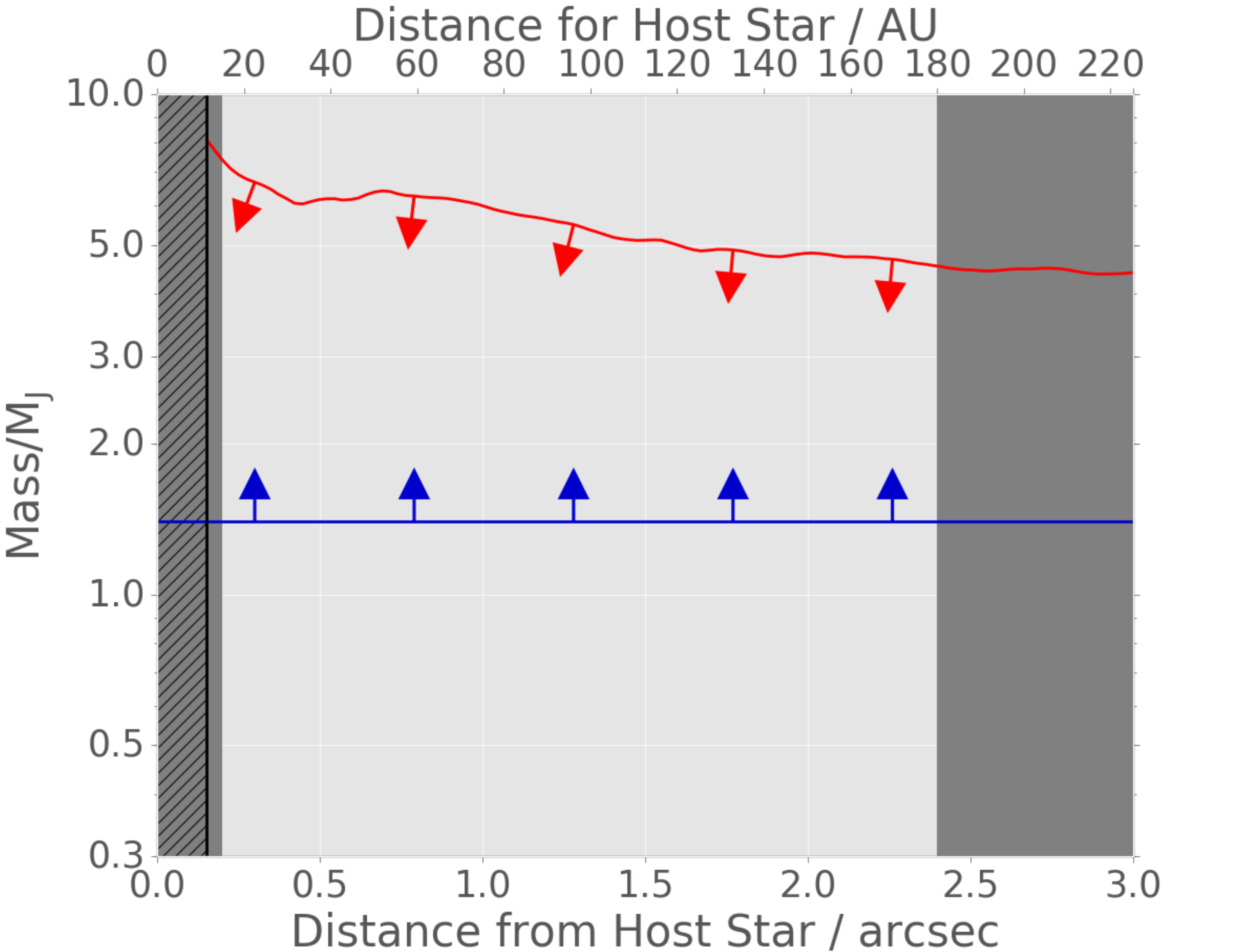}}
  %\subfloat
  \caption{Resultant knowledge of the planetary system around HD 38206, assuming an attempt at directly imaging planets with SPHERE results in a non-detection.  The non-detection gives a maximum planetary mass of 4.5 $m_J$~to 7.5 $m_J$~across the gap in the belt (red line), while the dynamical clearing constraint gives a minimum planetary mass of $\sim 1.4~m_{J}$~(blue line).  The gray areas are the locations of the debris, and the black striped area the inner working angle of SPHERE.}
\label{fig:sphereexample}
\end{figure}

New planet finding instruments such as GPI \citep{2014PNAS..11112661M} and SPHERE \citep{2008SPIE.7014E..18B} are able to image planets down to a few $m_J$.  For instance, an early paper on the 20 Myr system PZ Tel suggested a detection limit of $\sim$3$~m_J$ at 0.5" \citep{2016A&A...587A..56M}.  We consider the double debris disk star HD 38206, a 30 Myr old A0V star with a mass of $2.3~M_{\odot}$~\citep{1999A&AS..137..273G} at a distance of 75 parsecs.  HD 38206 was identified as a two temperature debris disk likely to contain two debris belts by \citet{2014MNRAS.444.3164K}.  Assuming blackbody grains, the debris rings are located at 15 au and 180 au.
We estimate the best contrast achievable by direct imaging as the 5-sigma contrast limit from unpublished SPHERE data for other systems (Matthews et al., in prep), by measuring the standard deviation in concentric annuli of the reduced image. This contrast is scaled to the distance and host magnitude for HD38206, and the detection limit is then converted to a mass limit using the COND models \citep{2003A&A...402..701B}.  Thus, if an observation of HD 38206 results in a non-detection of planets, the most massive planets that may be present would have $m_p \sim 5~m_{J}$, although it depends slightly on the separation from the star.  Using equation \ref{eq:minmass}, we calculate the minimum mass of planets needed to clear the gap in 30 Myrs to be $\sim 1.4~m_{J}$.  The mass limits correspond to three planets in the lower case, and two planets in the upper case.  We plot this example in figure \ref{fig:sphereexample}.  

Thus, by exploiting our knowledge of the debris disk, we are able to complement the upper mass limit derived from imaging with a lower mass limit derived from dynamics.  For this system this leads to an approximate knowledge of the planetary system, with the masses of the unseen planets known to less than an order of magnitude, and their number to $\pm1$.

\section{Discussion}
Implicit in this approach is the assumption that planets form surrounded by a sea of planetesimals that still retain a significant fraction of their mass.  There is some theoretical basis to believe planet formation may be $\sim 50\%$~efficient \citep{2004ApJ...614..497G,2004ARA&A..42..549G}.  There is some circumstantial observational evidence of this; the estimated mass of the Oort cloud \citep{2005ApJ...635.1348F,2015MNRAS.454.3267F}~and calculated fraction of small bodies that end up in the Oort cloud \citep[e.g.][]{2013Icar..225...40B,2015MNRAS.446.2059S}~imply the mass of solids scattered by the planets was comparable to the mass of solids in the planets.  Similar mass clouds may be commonly present around other stars \citep{2014MNRAS.445.4175V}.  Modelling of debris disks also suggests their total mass is comparable to the solid mass of planetary systems \citep{2008ARA&A..46..339W,2011ApJ...739...36S}.  The observational evidence does not strongly indicate that the proto-comets were co-spatial with the planets; if future observations fail to find the minimal planetary systems envisioned here, it will be significant evidence that planets do not clear gaps, but rather that planetesimal gaps form because planet formation is $\sim 100\%$~efficient, or that giant planets clear gas gaps that also removes solids \citep[as in][]{2016ApJ...825...77D}.  

Very recently, \citet{2016ApJ...823..118M} published a study on the maximal planetary system that can fit dynamically between two debris disks.  This provides a stronger constraint on older systems, and thus might provide a more stringent upper limit than direct imaging for older systems.

This model necessitates a caveat: we have neglected the mass of the planetesimals in our study.  If the mass of planetesimals is comparable to, or in excess of, that of the planets, they may cause migration of the planets \citep{1984Icar...58..109F}.  \citet{2014Icar..232..118M} published a set of criteria for when planets in a planetesimal disk may start to migrate.  If the minimum planetary system predicted by this study is such that migration might occur during the clearing phase, the model presented here may be inappropriate.  For the young systems most favourable to direct imaging, and most likely to host double debris disks, the minimum mass will be higher (equation \ref{eq:minmass}), and migration is unlikely to be a concern.  For instance, for HD 38206 we inferred at least $1500~m_{\oplus}$ in planets, while a typical A star debris disk is inferred to have a mass of $\sim 10~m_{\oplus}$ \citep{2007ApJ...663..365W}.  Consequently, from \citet{2014Icar..232..118M} we expect no migration, which only occurs for $m_p < 3~m_{disk}$.  A massive disk would also gravitationally self-excite, spreading the planetesimals \citep{1996Icar..123..180K}, and viscously spreading the small bodies \citep{2013A&A...558A.121K}.  This could allow them to encounter secular resonances and be cleared on shorter timescales.  As the spreading will depend on the mass and size distribution in the debris, there is no good way to estimate the appropriate timescale.

\section{Summary}
We present a simple equation for the minimum mass of planet needed to clear the gap in double debris disk systems (equation \ref{eq:minmass}), and the number of such planets that would typically be found in the gap (equation \ref{eq:numberofplanets}).  At least one direct imaging survey \citep{2015ApJ...800....5M}~has begun targetting double debris disks to search for planets.  Currently, if no planets are detected, we can only infer that planets with masses less than the detection limit may lie within the gap.  By imposing constraints on both the minimum and maximum planetary systems that could be present, the observational non-detection of planet(s) can be recast as more positive knowledge about the planetary system harboured by the star in question.  The use of the clearing time provides the strongest constraints on young systems, as does direct imaging \citep{2003A&A...402..701B}, providing a natural synergy.

\section{Acknowledgements}

Andrew, Amy, and Quentin are supported by the European Union through ERC grant number 279973.  We thank Sasha Hinkley for prompting us to consider the problem, Grant Kennedy for useful discussions, and the anonymous referee whose comments led to an improved manuscript.

\bibliographystyle{mn2e}
\bibliography{clearing}

\end{document}